\documentclass{article} 
\usepackage{nips15submit_e,times}
\usepackage{hyperref}
\usepackage{url}
\usepackage{graphicx}
\usepackage{balance}
\usepackage{comment}
\usepackage{balance}
\usepackage{comment}

\usepackage{subfigure}
\usepackage{algorithmic}
\usepackage{algorithm}
\usepackage{verbatim}
\usepackage{subfigure}
\usepackage{algorithmic}
\usepackage{algorithm}
\usepackage{verbatim}

\title{ Finding Nearest Neighbors in  graphs locally }

\author{
Abhinav Mishra \\
College of Computing \\
Georgia Institute of Technology \\
\texttt{amishra41@gatech.edu} \\
}

\nipsfinalcopy 

\begin{document}

\maketitle

\begin{abstract}
Many distributed learning techniques have
been motivated by the increasing size of datasets and their inability to fit
into main memory on a single machine. We propose an algorithm that finds the nearest neighbor in a graph locally without the need of visiting the
whole graph. Our algorithm is distributed which further encourage scalability.
We prove the convergence of the algorithm.\\\
\end{abstract}

\section{Introduction}

Graph based learning methods have applications in spectral algorithms [3], dimensionality reduction[9], image-segmentation[2], semi-supervised learning [8], manifold learning [10],
 link prediction [4], databases[5] etc. Some of the desirable properties of any graph based algorithms include scalability where we can run the algorithm on networks such as Facebook involving billions of nodes, ability to handle dynamic environment (addition/removal of nodes or edges), online computation (avoiding pre-computation such as matrix inversion),  ability to run in a distributed setting where a node communicates with its neighbors.

Finding nearest neighbors in a graph is one of the fundamental problem and is applied to many applications mentioned above. 
 Nearest neighbors are the nodes that are close to a given node based on a certain proximity measure. We expect 
a proximity measure to be able to capture the structure of the graph, i.e., if two nodes have multiple paths between them and fewer hops between them,
then the nodes are said to be  in a close proximity.   We discuss the properties of proximity measures in detail in section 2. 
Traditionally, random walk based methods are used in identifying the nearest neighbors because its ability to look beyond one hop and capture the network structure. Random walk methods can be further divided into two types: probability of hitting a node also usually known as Personalized-pagerank (PPR)  and expected number of steps to reach a node (hitting time or commute time) [6].

While it is exciting that a method such as random walk captures the relationship between two nodes in a good way, i.e., it favors the multiple short-paths between two nodes. However, random suffers from a critical problem. For finding nearest-neighbors, we  require algorithm to capture the local structure instead of the whole graph.  For example, consider two dense components connected by few edges. We expect a node and
its neighbors to lie in the same component. Therefore,  adding more edges or making other changes in another component should not have any significant effect on a good proximity measure. 
However,  a personalized pagerank will be reduced by roughly one-fourth if we induce extra $|E|/3$ on a node from another component or add similar edges. To be more precise, in an undirected graph, the steady state distribution of a node is proportional to its degree in the graph. Such graphs are common, e.g., consider dense giant components connected with few edges.
Any random walk based proximity measure is likely to be effected by the presence of high degree nodes. Similar results were drawn for hitting and commute 
times[1]. A good  local proximity measure relies on the fact that a node should not be effected by any  node that is far away. There is a line of work that focus on making random walk local 
of work such as truncated time. 
Second issue with random walk based approaches is the computational load. Computing commute time typically involves finding a pseudo-inverse on laplacian which  is $O(n^3))$. Spielman made it finding in nearly linear time. However, if the graphs are changing, then even such approach is computationally expensive. At the same time, finding an algorithm in a distributed setting is of interest because it does not require any pre-computation and is naturally scalable.

These issue gives rise to few fundamental questions. 1) can we derive a measure that is truly local, i.e., complexity is independent of number of nodes in the graph, measure to obey the properties such as favoring short-multiple paths between the nodes and captures the local-structure of the graph.
The  networks such as facebook are dynamic, where  node or edges can be added or deleted. This give rise to a second important question Property 2: can we
avoid any kind of  pre-computation on the graph such as matrix inversion, or. Therefore, if there are any changes. on the fly computation.    Recently there is a large amount of research on scalability and storage due to the increasing size of datasets and  inability to fit into main memory on a single machine.
So, this leads to the final question of devising an algorithm that is truly distributed in nature. We assume the simplest setting where a node in graph is allowed to interact with its neighbors.

We propose an iterative algorithm that runs on a graph under distributed setting where a node  contact its neighbors. 
Initially, the querying node has a finite charge and it starts spreading the 
charge to its neighbors and neighbors repeat the process and so on. Each node retain a certain amount of charge and therefore, we can expect fewer 
nodes to have charge and charge cannot reach the nodes that are far away. Our algorithm is a non-linear process but has s resemblance with Markov Chains, It terminates after few steps, instead of converging asymptotically. We prove the termination and present an upper-bound on the number of steps.
 Our algorithm
runs on small part of the graph which is controlled by the user. Hence, the complexity of the algorithm is independent of the number 
of nodes. This makes the algorithm desirable for large graphs, where we can avoid the requirement to visit all the nodes. Algorithms
such as Personalized Pagerank, commute time based measures. other global are global and do require the whole graph. It may need the whole Adjacency representation at once
for the purpose of inversion, or may require all the nodes by .  However, our algorithm explores a smaller part of the graph to find the $k$-Nearest Neighbors.

\begin{figure}
\centering
\includegraphics[width=0.4\textwidth]{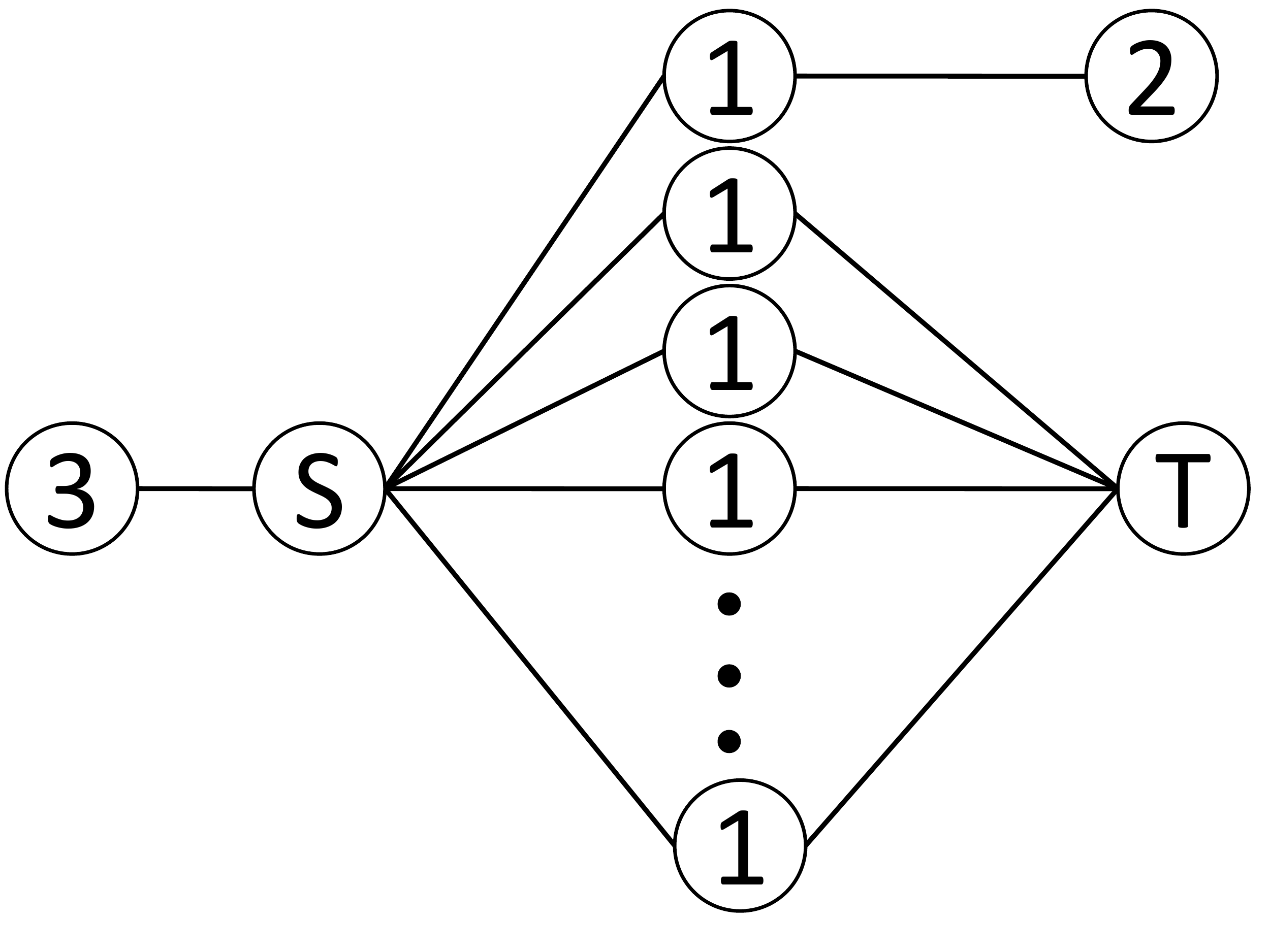}
\caption{Left subgraph of the starting node has a fewer edges and therefore has more active nodes as the charge is equally divided between edges. Right subgraph is dense and has fewer active nodes. }
\label{corevsperi}
\end{figure}

\section{Pursuit for Proximity measure }
The nodes within a close proximity are expected to be a good candidate for Nearest Neighbors. There are number of  factors that  effect proximity. Simplest measure is the 
number of hops between two nodes. If the shortest path between two nodes is large, then it is expected that they are not close neighbors.
However, this shortest path does not factor the multiplicity of paths. For example,
if there are multiple paths between two nodes, it indicates that the two nodes might be well connected. As shown in Figure 1, node S (Source) is 
connected to node T (sink) through multiple paths compared to node 2 which has only one path from node S. Notice that both paths from S
to T and S to node 2 are of length 2 each. However, since node T has more paths  from S, it will  have closer proximity compared to Node 2. So far the
the two proximity measures (namely shortest path and number of paths) we have seen are symmetric, i.e., we obtain the same score from both source and destination.  Now we introduce the third measure that brings asymmetricity. It is not difficult to see that Nearest Neighbors (NN) in general is not a symmetric measure.
It is not necessary for a node say A to be a NN of node B, given that node B is in a NN of A. 

 A node with a high degree is less likely to have 
one of its neighbors as NN compared to a node with a low degree. However, a good measure of proximity should handle all three cases together.
In isolation, 
For example in fig, 
We can draw parallel of our methods with Markov Chains in all three scenarios mentioned above. 

Commute time or hitting time essentially captures all the three characteristics. These two measures are also related to the electrical resistances
between two nodes in a graph, where we treat edge weights as resistances. However, there are few problems with this measure. First is the computation.
One generally require the whole graph for computing the commute time or hitting time. Therefore, the computational complexity depends on the number
of nodes or edges. Secondly, even if the graph is stored with a succent representation for calculating the such measure. It suffers from the dynamic 
nature , where a nodes or edges are added or deleted and therefore, creates an extra overhead of creating a sub-structure. 
In Markov Chains, it can be seen as 
If node is removed then the resulting matrix will be sub-stochastic. We later show that by setting appropriate parameters, our algorithms becomes
a Markov Chain. As discussed earlier, the technique we propose here is a non-linear iterative method, and by reducing the non-linearity, we reduce it
to a Markov Chain. However, such transition from non-linearity to linearity has a lot of implications. Firstly, due to non-linearity, we only explore a
small part of graph and secondly, the algorithm terminates after finite number of steps, while the markov chains converge asymptotically.

\section{Algorithm}
\label{sec:algo}
Consider $G(V,E)$ as a connected undirected graph, $|V| \! = \! n$,
$d_i$ is the degree of node $i$ and $d_{\max} \! = \! \max \{ d_i : i \in V \}$, and
$\vec{x^0} = (x_1^0 , \ldots , x_n^0)$ is a non-negative function over the set of nodes: $x_i^0 \! \geq \! 0$. 
We call $x_k^i$ as the charge of node $i$ at iteration $k$. We assume that the charge is non-negative for any node at any iteration.
We also put a constraint that $\sum_{i \in V} x_i^0 = 1$.  We further assume that there is only one node in the beginning with a unit charge.
Our goal is to find $k$-nearest neighbor with respect to this node.

We assume that the whole network of nodes is distributed, i.e., there is not any central node and computation  at each node are
independent and synchronous. For a node to communicate with a node 3-hops away, it has to do it through the neighbors. -advantages-

We technique is an iterative method, where at each iteration  nodes distribute the charge among themselves. The sum of the charge remains 
the same in each iteration. Also, there is a charge conservation at each node.  We start with a single node with a unit charge, then the node spreads
the charge to its neighbors and the neighbors so the same in the next iteration and so on. Therefore, with more iterations, the a subgraph induced by
the charged nodes start spreading. Nodes that receive charge retain a certain amount all the time. Therefore the amount of charge that can be 
 spread decreases with the increase in charged nodes. The  amount of charge every charged node retains is
between $\left[(1-\alpha)\cdot \epsilon, \epsilon\right)$. Therefore, there are limited number of nodes that can spread the charge due to the conservation
of charge at each iteration. By choosing the suitable $\epsilon$ and $\alpha$, we can control the size of subgraph formed by the charged nodes. We set formed by the charged nodes, we call as $k$-NN set.
We now formally give the details of the algorithm.

For a node $i$ at iteration $t$, we define a variable $z_i^t$ such that

\begin{displaymath}
z_i^t \,\,= \,\, \left\{ 
\begin{array}{lll}
1 & {\rm if} & x_i^t > \epsilon \\
0 & {\rm if} & x_i^t \leq \epsilon
\end{array}
\right. 
\end{displaymath}

\noindent
A node $i$ with $z_i^t=1$ is known as an active node in iteration $t$ and are the only nodes that are allowed to transfer the charge.
Other nodes can only receive the charge from active nodes. Therefore,  nodes with a charge more than $\epsilon$ can transfer the charge. 
For a node $i$ at iteration $t$, the amount of charge it carry in iteration
$t$ is represented by $x_i^{t}$, and in $t+1$ iteration by $x_i^{t+1}$. Let $d_j$ be the degree of node $j$. The charge $x_i^{t+1}$ is computed as follows:

\begin{equation}
\label{p1}
x_i^{t+1} \,\,= \,\,\left( (1-\alpha) x_i^t z_i^t + x_i^t (1-z_i^t) \right) \,\,+ \,\,
\alpha \sum_{j : \{ i,j \} \in E } \frac{x_j^tz_j^t}{d_j} 
\end{equation}

\noindent

The first part of the equation $\left( (1-\alpha) x_i^t z_i^t + x_i^t (1-z_i^t) \right)$ indicates the amount of charge node $x_i^{t+1}$ retains.
If $z_i^t=0$, then the node retains all the charge it had in the previous iteration, and therefore by the conservation of charge at the node , it
cannot send charge to its neighbors. If $z_i^t=1$, then the node retains $(1-\alpha)$ fraction of the charge as given by $(1-\alpha) x_i^t z_i^t $, and it sends rest of the
charge to its neighbors.

Second part of the equation $ \alpha\sum_{j : \{ i,j \} \in E } \frac{x_j^tz_j^t}{d_j} $ indicates the amount a node receives from its neighbors.
Since only the node, with $z^t_*=1$ are allowed to transfer the charge, therefore we have an indicator variable $z^t_*$ indicating if the node is eligible to send the 
charge. Since the nodes with $z^t_*=1$ retain the $(1-\alpha)$ fraction of the charge, they  distribute the other $\alpha$ fraction to its neighbors. In our process, they distribute
it equally to their $d_j$ neighbors. Here $d_j$ indicates the degree of node $j$. It is not difficult to verify the charge conservation with
the above formulation.

The above formulation is a non-linear system of equation that terminates 
after finite steps using fixed-point iterations (proved in ).  A necessary condition for termination is 
This non-linearity makes it 
difficult to analyze. We present a upper-bound on the number of steps required for termination. 
We discuss the running time in Section~\ref{sec:complex}. We prove the termination of our
algorithm in Section~\ref{sec:term} and also give an upper-bound on the number of iterations required
for a guaranteed termination. The algorithm exhibits a locality property that
the charge is disseminated to the nodes that are closer in proximity. We discuss this locality in detail
with applications in Section~\ref{sec:local}.

\section{Discussion}
The main intuition behind finding the NN is that if a node has a charge and it distribute it, then it will reach its NN first because of the multiple short paths
between them. Next challenge is to formulate an algorithm that is scalable, terminates/converges, scalable and distributed. It is easy to see that 
the maximum number of $k$-NN nodes is $\frac{1} {(1-\alpha)\cdot\epsilon}$, otherwise the net charge will be more than 1. We give various properties and
their proof in Appendix.

The necessary condition for termination in iteration $t$ is $x_i^{t} \leq \epsilon, \forall i$, since if there is not a single active node, then there
is no exchange of charge, so the process terminates. Once a nodes become active, the minimum charge it will retain is $(1-\alpha)\cdot \epsilon$. Here,
$\alpha$ is a user controllable parameter and plays an important role in termination and setting an upper-bound on the number of steps
required for termination. Let the set of  nodes that were once active be called as NN set. More formally, NN$=\{i \in V: \{x^t_i \geq \epsilon, \forall t \geq 0\}\}$.   

\subsection{Properties of NN subgraph}
The maximum number of $k$-NN nodes is $\frac{1} {(1-\alpha)\cdot\epsilon}$, otherwise the net charge will be more than 1 which is not possible. Moreover, the subgraph induced by the $k$-NN nodes is well-connected. It follows from the fact that
a node  can only  receive charge from an active node and once a node becomes active, it belongs to $k$-NN set.
The subgraph induced by the $k$-NN nodes is well-connected. It follows from the fact that
a node  can only  receive charge from an active node and once a node becomes active, it belongs to $k$-NN set.



\section{Complexity and  termination}
\label{sec:term}
{\bf Proof of termination.}
We start with an important observation that
there is always a charge that either moves out of core or moves towards the periphery (while staying in the core). 
We quantify the charge later in the section. For now, say $\beta$ is the minimum charge that continues to move out of
core say in at most $t$ iterations, then we  conclude that after $t/\beta$ iterations, the algorithm will terminate. 
Likewise if $\beta$ is the minimum charge that moves a step closer to the periphery every $t$ iterations, then it will
take $t/\beta$ iteration to move a unit charge a step closer to the periphery. If the diameter of core is $d$, then it will
take $d\cdot t/\beta$ iterations for the charge to move out of core. The maximum amount of charge a periphery can hold is
$1- (1-\alpha)\cdot \epsilon$ (for a star graph). Therefore, it will take $(1- (1-\alpha)\cdot \epsilon)d\cdot t/\beta$ iterations at most.
In case of star graph, it is a single iteration. But the above approach is more systematic and is applicable to any graph.

Note that the bound given above is really loose, as the nodes in core will retain at least $(1-\alpha)\epsilon$ charge. But it does
prove that as long as there is a fixed amount of charge coming out of core every few iterations, the process will terminate. Secondly,
it is a progress if a charge moves towards the periphery while still residing inside the core, as it will eventually move out of core. 

We now quantify the values such $\beta$ mentioned above. There could be at most $1/(1-\alpha)\epsilon)$ core nodes (property vi). Therefore,
the diameter of the core is at most $1/(1-\alpha)\epsilon$, otherwise the $\sum_ix_i>1$, which is not possible. The minimum charge a node can give out is at least $\frac{\alpha\cdot \epsilon}{d_{max}}$. Therefore, it will take $\frac{d_{max}}{\alpha\cdot(1-\alpha)\cdot \epsilon^2}$ steps to pass at most the unit
amount of charge. 

This bound is still loose. We further improve it by introducing the concept of excess charge. Excess charge is defined as the maximum
amount of charge a core can give out to periphery. For example, if there are $M$ nodes in core, than $1- (1-\alpha)\cdot \epsilon\cdot M \leq1- (1-\alpha)\cdot \epsilon$ is the amount of excess charge, as each core node will retain at least $(1-\alpha) \cdot \epsilon$ charge. Plugging the excess charge into the earlier bound, we obtain an improved
bound of $(1-(1-\alpha)\cdot \epsilon)\cdot \frac{d_{max}}{\alpha\cdot \epsilon}\cdot M \leq (1-(1-\alpha)\cdot \epsilon)\cdot \frac{d_{max}}{\alpha\cdot (1-\alpha)\epsilon^2}$.

{\bf Complexity and Upper bound on the number of steps}
In any iteration, a node communicates with its neighbors. Therefore, the exact number of communication exchange is $\sum_{i \in V} d_i=2|E|$. If the algorithm 
terminates in $k$ steps , the complexity becomes $O(k|E|)$. However, it is usually the case that algorithm runs on a small part of graph. The minimum charge a peripheral node can have is $\frac{\alpha\cdot \epsilon}{d_{max}}$ other than the nodes with zero charge. So there are at most $\frac{d_{max}}{\alpha\cdot \epsilon}$ number
of nodes with a non zero charge. Hence the complexity now becomes $O(k\cdot \frac{d_{max}^2}{\alpha\cdot \epsilon})$. 

\section{Weighted Graphs and Markov Chains}
It is straightforward to extend the formulation to the directed-weighted graphs 
if the graph is  strongly connected. For weighted-undirected graph,
assume that there exist a directed edge in both direction. Since the algorithm works on a small part of graph, assume that $(1-\alpha)\epsilon << n$.
Assumption of strong connectivity is necessary, although not in every case. Suppose that we start with a node that does not have any outgoing edges,
then the node cannot transfer the charge. We do not require the condition of aperiodicity, as . This is another difference with markov chains.

Let $w_{jk}$ denote the edge weight of an edge going from node $j$ to node $k$. 

\begin{equation}
\label{p1}
x_i^{t+1} \,\,= \,\,\left( (1-\alpha) x_i^t z_i^t + x_i^t (1-z_i^t) \right) \,\,+ \,\,
\alpha \sum_{j : \{ j\rightarrow i \} \in E } \frac{x_j^tz_j^t}{\sum_k w_{jk}} 
\end{equation}

In this section, we show that our method is similar to markov chain and in fact, it becomes a markov process if we set $\epsilon$ to zero and enforce that $\sum_{i \in V} x_i^0 = 1$, so that it becomes a probability distribution. 
To see this, $z_*^*$ becomes 1 for all nodes, and the equation becomes

\begin{equation}
\label{process1}
x_i^{t+1} \,\,= \,\,\frac{1}{2} x_i^t \,\,+ \,\,
\frac{1}{2}\sum_{j : \{ i,j \} \in E } \frac{x_j^t}{d_j} 
\end{equation}

The above equation represents lazy random walk, where a node retains half of the charge each time and distributes the other half. In random
walk terms, random walk stay at the same node with probability $0.5$ and moves to one of its neighbor with probability $1/2d$. Lazy random walk are 
typically used to fix up the periodicity problem. For example, in a bipartite graph, a walk will be either on the one side or another, and thus fail to converge.
Self loop also helps in convergence rates as it bounds the eigenvalue of the transition matrix. We can observe the similar phenomenon with our formulation,
the self absorbing parameter $\alpha$ and $\epsilon$ give an upper-bound on the number of steps required for termination.
qualitative treatment with random walk

\section{ Terminating iterative method}

It remains an open question to show the convergence rate of the earlier technique.
We now present an slightly modified version of the earlier algorithm. Most the properties remain the same. In this technique, a node looses 
$(1-\alpha)$ fraction of access charge at each iteration. In other words,  it has charge $\alpha (x_i-\epsilon)$  to distribute to its neighbors.
If the charge is less than $\epsilon$ then the node retains most of its charge. So the modified method becomes
\begin{equation}
\label{p1}
x_i^{t+1} \,\,= \,\,\left(  x_i^t z_i^t + x_i^t (1-z_i^t) \right) \,\,+ \,\,
\alpha \sum_{j : \{ j\rightarrow i \} \in E } \frac{x_j^t(1-z_j^t)}{ d_j} 
\end{equation}

To see the convergence rate, we can write the formulation as 
\begin{equation}
\label{p1}
x_i^{t+1} \,\,= \,\,\left(  x_i^t z_i^t + x_i^t (1-z_i^t) \right) \,\,+ \,\,
\alpha \sum_{j : \{ j\rightarrow i \} \in E } \frac{x_j^t(1-z_j^t)}{ d_j} 
\end{equation}

Let $(a^1,... a^t)$ be the excess charge in the graph. We define excess charge as $a^t=\sum_i (x^t_i-\epsilon)(1-z^t_i)$.
Notice that $1 \geq a^i \geq a^j, \forall i \leq j$. In every iteration, excess charge is reduced by fraction $\alpha$.
Therefore after $t$ iterations, excess charge $a^t$ is at most $\alpha^t$ after $t$ iteration. Since excess charge is what is 
transferred between core nodes and peripheral nodes, for convergence we need it be $a^t < \delta$. Therefore, we need $log \delta /\log \alpha$ iterations.

\begin{equation}
x_i^{t+1}= 
 \frac{ x_i^t+ \epsilon}{2}+
\frac{1}{2d} \sum_{j : \{ i,j \} \in E,j \in C } {(x_j^t-\epsilon} ) \\
= \frac{1}{2d} \sum_{j : \{ i,j \} \in E,j \in C } {(x_j^t+x^t_i} )
\end{equation}

Now we look at the vector $|| \vec{x}^{t+1}||= \sum_{j \in C} (x^{t+1}_j)^2$, and  obtain an upper-bound.
\begin{equation}
|| \vec{x}^{t+1}||= \sum_{j \in C} (x^{t+1}_j)^2\\
= \sum_i \left (\frac{1}{2d} \sum_{j : \{ i,j \} \in E,j \in C } {(x_j^t+x^t_i} ) \right)^2\\
\end{equation}

\begin{equation}
\leq \sum_i \frac{1}{d} \sum_{j : \{ i,j \} \in E,j \in C } \frac{{(x_j^t+x^t_i} )}{4} ^2\\
= \frac{1}{d} \sum_{{ i,j } \in E,{i,j} \in C } \frac{{(x_j^t+x^t_i} )}{4} ^2
\end{equation}

Since all the core nodes have at least $\epsilon$ charge,
$a^t-a^{t+1}$, the amount peripheral nodes receive in iteration $t+1$.

\section{Related work}

Different measure for proximity have been studied in detail in [4]. It was observed that classical measure such as katz outperform more appealing methods such as commute time especially in the presence of a high-degree node[7]. Sarkar also observed the global nature of the walk and proposed
methods based on the fixed length walk and called it truncated hitting time. Here the paths of length less than $T$ are considered. The total running
time was $O(n|E|)$, here $n$ is the number of nodes, $|E|$ is the number of nodes. One first problem is the time complexity and other one is the algorithm
does not work on a dynamic graph. To overcome this, authors later refined their hitting time computation mechanism by sampling. Note that the computation
is now approximately correct with a hight probability instead of deterministic in the earlier work. This way they reduced the complexity to to $O \sqrt(n)$
for a pair which is very reasonable.

Knowing $T$ apriori is a challenge. However, one can argue that since $T$ is a constant, one can perform a linear search on the multiple values of $T$.
But there is a major problem with such approach,  graph such as Facebook where average distance is 4.2 or classical milligram experiment[4] . Choosing 
$T=3$ and $T=4$ will capture entirely different graph. This seem to indicate that perhaps a fixed length walk is not the most suitable method.


\section*{APPENDIX}
\subsection*{\bf A.   Properties of the Algorithm}
For a subset of nodes $H \! \subset \! V$, we use $\Gamma_1 (H)$ to denote
the 1-vertex neighborhood of $H$: 
$\Gamma_1(H) = \{ j \! \in \! V \setminus H : 
\{ i^\prime \! \in \! H : \{ i^\prime , j \} \! \in \! E \} \! \neq \! \emptyset \}$.
Let $i_0 \! \in \! V$ be a node such that $x_{i_0}^0 \! = \! 1$ (consequently $x_i^0 \! = \! 0$, 
$\forall \! i \! \neq \! i_0$) and let $n \geq\frac{1} {(1-\alpha)\cdot\epsilon}$. 
Then following properties are true:

$~~~~${\bf (i)  {\boldmath${ \epsilon}\cdot (1-\alpha) < x_i^* \leq 1$, $\forall i \in H$.}} \
 If a node has a charge more than $\epsilon$, it will retain at least $\epsilon\cdot (1-\alpha)$ of the charge (by Eq. 1). All the other nodes with charge more than $\epsilon\cdot (1-\alpha)$, but less than $1$, will continue to hold all of the charge and will not distribute it to their neighbors.  \\
$~~~~${\bf (ii) The subgraph of {\boldmath$G$ induced by \boldmath$H$ is connected.}} 
 A node can only receive a charge
from a core node. Therefore there has to be a path of core nodes between a node receiving a charge and $i_0$ (starting node). Also, a node that becomes
a core node continues to be a core node, proving the connectivity. \\
$~~~~${\bf (iii) {\boldmath$x_i^{t+1} = x_i^t$, \boldmath$\forall i \in V$ and \boldmath$\forall t \geq t_0$. }}
 See Section \ref{sec:term} \\
$~~~~${\bf (iv)} {\boldmath$0 \leq x_i^t < \epsilon\cdot (1-\alpha)$, $\forall i \in \Gamma_1 (H)$. }
Here, $ \Gamma_1(H) $indicates the set
of nodes that are neighbor of core including the core nodes itself. $ H \cup \Gamma_1(H)$ shows the peripheral nodes. Such peripheral nodes can
receive the charge from core since they are the neighbors of core, but their charge has to be less than $\epsilon\cdot (1-\alpha)$, otherwise they become core.\\
$~~~~${\bf (v)} { \boldmath $x_i^t = 0$, $\forall i \in V\setminus\left( H \cup \Gamma_1(H) \right)$
and $\forall t \geq 0$.} All the nodes that neighbors of peripheral nodes, but not a part of core will have zero charge, as the peripheral node can have 
at most $\epsilon\cdot (1-\alpha)$ charge and are not allowed to distribute it to their neighbors.\\
$~~~~${\bf (vi)}  { \boldmath$|H|\leq 1/\epsilon\cdot (1-\alpha)$.} 
The minimum charge a core node can have is $\epsilon\cdot (1-\alpha)$ and 
charge of a starting node $1$, there by the conservation of charge, there could be at most $1/(\epsilon\cdot (1-\alpha))$ core nodes.\\
$~~~~${\bf \boldmath (vii) $|V|\leq 1/\epsilon$, then algorithm does not terminate. }
 If the size of the graph is less than $1/\epsilon$, then there is at least one node with charge more than $\epsilon$ and will continue to give charge to
its neighbors, therefore the process cannot terminate.

\section*{References}

\small{
[1] U. von Luxburg, A. Radl, and M. Hein. Hitting and
commute times in large random neighborhood graphs.
Journal of Machine Learning Research, 15:1751–
1798, 2014.

[2]  J. Shi and J. Malik. Normalized cuts and image segmentation.
IEEE Transactions on Pattern Analysis
and Machine Intelligence, 22(8):888–905, 2000.

[3]  F. Chung. Spectral Graph Theory. American Mathematical
Society, 1997.

[4] D. Liben-Nowell and J. Kleinberg. The link prediction problem for social networks. In CIKM, 2003.

[5] Balmin, A., Hristidis, V., and Papakonstantinou, Y. (2004). ObjectRank: Authority-based keyword search in databases. VLDB, 2004.

[6] Aldous, D., and Fill, J. A. (2001). Reversible markov chains.

[7] Sarkar, P., and Moore, A. (2007). A tractable approach
to ¯finding closest truncated-commute-time neighbors in
large graphs. Proc. UAI.

[8] Zhu, X., Ghahramani, Z. and Lafferty, J. (2003) Semi-Supervised Learning Using Gaussian Fields and
Harmonic Functions. ICML 20: 912–919.

[9]J. B. Tenenbaum, V. de Silva, and J. C. Langford. A global geometric framework for nonlinear
dimensionality reduction. Science, 290:2319–2323, 2000.

[10]M. Belkin and P. Niyogi. Laplacian eigenmaps for dimensionality reduction and data representation.
Neural Computation, 15(6):1373–1396, 2003.
}
\end{document}